\documentclass[prl,twocolumn,superscriptaddress]{revtex4-1}
\usepackage{graphicx}
\usepackage{dcolumn}
\usepackage{bm}
\usepackage{color}
\usepackage{amsmath}
\usepackage{amsfonts}
\usepackage{soul}

\begin{document}

\title{Topology Restricts Quasidegeneracy in Sheared Square Colloidal Ice}

\author{Erdal C. O\u{g}uz}
\affiliation{School of Mechanical Engineering, Tel Aviv University, Tel Aviv 69978, Israel}
\affiliation{School of Chemistry, Tel Aviv University, Tel Aviv 69978, Israel}
\affiliation{Sackler Center for Computational Molecular and Materials Science, Tel Aviv University, Tel Aviv 69978, Israel}

\author{Antonio Ortiz-Ambriz}
\affiliation{Departament de F\'isica de la Mat\`eria Condensada, Universitat de Barcelona, Barcelona, Spain}
\affiliation{Institut de Nanoci\`{e}ncia i Nanotecnologia, Universitat de Barcelona, Barcelona, Spain.}

\author{Hadas Shem-Tov}
\affiliation{School of Physics and Astronomy, Tel Aviv University, Tel Aviv 69978, Israel}

\author{Eric Babi\`a-Soler}
\affiliation{Departament de F\'isica de la Mat\`eria Condensada, Universitat de Barcelona, Barcelona, Spain}

\author{Pietro Tierno}
\affiliation{Departament de F\'isica de la Mat\`eria Condensada, Universitat de Barcelona, Barcelona, Spain}
\affiliation{Institut de Nanoci\`{e}ncia i Nanotecnologia, Universitat de Barcelona, Barcelona, Spain.}
\affiliation{Universitat de Barcelona Institute of Complex Systems (UBICS), Universitat de Barcelona, Barcelona, Spain.}

\author{Yair Shokef}
\email{shokef@tau.ac.il}
\homepage{https://shokef.tau.ac.il}
\affiliation{School of Mechanical Engineering, Tel Aviv University, Tel Aviv 69978, Israel}
\affiliation{Sackler Center for Computational Molecular and Materials Science, Tel Aviv University, Tel Aviv 69978, Israel}
\affiliation{Center for Nonlinear Studies, Los Alamos National Laboratory, Los Alamos, NM 87545, USA}

\begin{abstract}
Recovery of ground-state degeneracy in two-dimensional square ice is a significant challenge in the field of geometric frustration with far-reaching fundamental implications, such as realization of vertex models and understanding the effect of dimensionality reduction. We combine experiments, theory, and numerical simulations to demonstrate that sheared square colloidal ice partially recovers the ground-state degeneracy for a wide range of field strengths and lattice shear angles. Our method opens an avenue to engineer a novel class of frustrated micro- and nano-structures based on sheared magnetic lattices in a wide range of soft- and condensed-matter systems.
\end{abstract}

\maketitle

\begin{figure}[t]
\includegraphics[width=0.98\columnwidth]{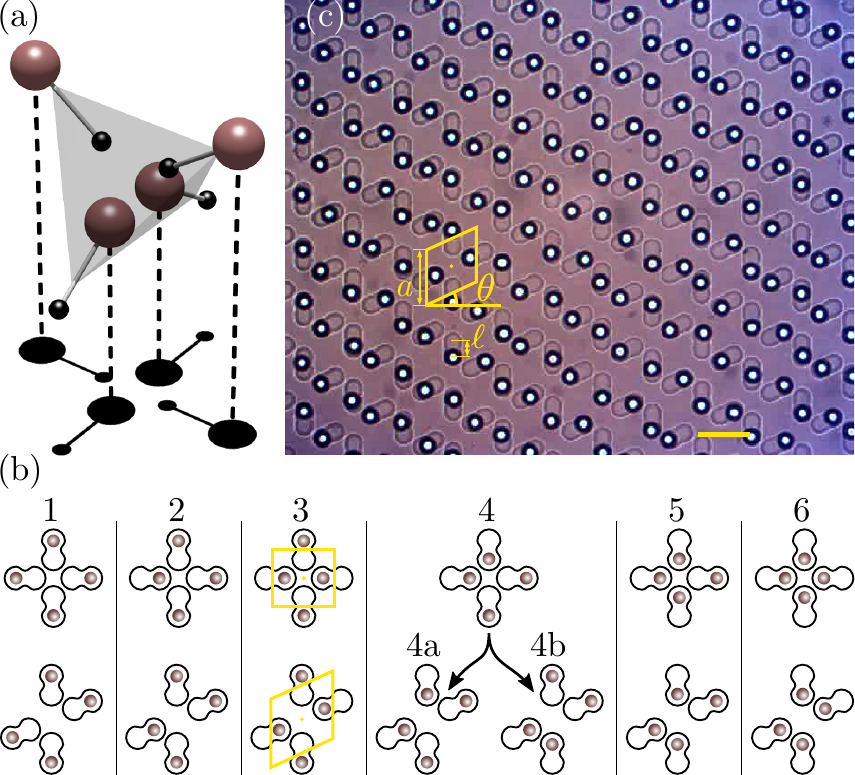}
\caption{(a) Schematic 
of one of the lowest energy configuration of water ice I$_h$ (top) and the corresponding projection on a 2D plane to give the square lattice. (b) In 2D square colloidal ice (top), vertex type 3 has lower energy than type 4, in which the two in-particles are closer to each other. Upon shearing the lattice (bottom), type 4 splits so that the energy of type 4a approaches that of type 3, as the shear angle $\theta$ increases. (c) Experimental realization of a sheared square colloidal ice with lattice constant \textit{a}, trap length $\ell$, and shear angle $\theta=25^\circ$ under a magnetic field $B=7.9$mT. Scale bar $30 \rm{\mu m}$. }
\label{fig:exp_and_types}
\end{figure}

The extensive entropy of ice at zero temperature has been elegantly explained as the result of the degeneracy of its energy-minimizing local atomic configurations~\cite{Pauling1935}. In hexagonal ice (I$_h$), the Oxygen ions form a tetrahedral network 
of hydrogen bonds, and there are six possible 2-in-2-out configurations, obeying the so-called ice rule, where two protons are near
an Oxygen, and two away from it, see Fig.~\ref{fig:exp_and_types}(a). Since the distances between the protons are the same at each vertex in this three-dimensional (3D) structure, these configurations have the same energy, making the ground state degenerate.

Recently, artificial mesoscopic systems, comprised of arrays of building blocks that are macroscopic enough to be directly visualized, such as artificial spin ice~\cite{Wang2006, NisoliRMP2013, Skjaervo2019}, buckled colloidal monolayers~\cite{Han2008, Shokef2011}, and mechanical metamaterials~\cite{Kang2014, Coulais2016, Meeussen2019a, Meeussen2019b} have been valuable in gaining insight on geometric frustration, and on the resulting topological features in ice-like systems. However, most of these systems are based on two-dimensional (2D) or quasi-2D lattices and, while enabling direct visualization of the relevant degrees of freedom, they introduce geometric constraints which profoundly influence the system's frustration. This particularly affects the 2D square geometry, which relates to a projection of the 3D ice on a plane, see Fig.~\ref{fig:exp_and_types}(a), and where the distances between the elements at each vertex are not the same. Thus, the corresponding ground state becomes non degenerate. As a consequence, these 2D structures  cannot be used as a physical realization of vertex-type models, the latter being characterized by extensive ground state degeneracy~\cite{Lieb1967}. Such models are cornerstone in statistical mechanics, and predict exotic behavior with algebraic correlations, extensive degeneracy, and direct connection to gauge theories~\cite{BaezBook} and quantum systems~\cite{EckleBook}. 

Recent attempts aimed at restoring the degeneracy of 3D water ice in 2D artificial-spin-ice systems using out-of-plane offset~\cite{Moller2006, Perrin2016, May2019}, by manipulating the magnetic interactions between elements in the array~\cite{Ostman2018, Perrin2019, Caravelli2019}, and by using other symmetries in the 2D plane~\cite{Tanaka2006, Morrison2013, Chern2013, Gilbert2014, Macedo2018, Macauley2019}. An alternative class of artificially frustrated systems which can resemble water ice, is the colloidal ice, where interacting microscale particles are confined in a lattice of double well traps~\cite{Libal2006,Ortiz2019}. Similar to artificial spin ice, in a square colloidal ice the repulsion between the particles gives rise to a 2-in-2-out ice rule for the four traps meeting at every vertex of the lattice~\cite{Libal2006, Ortiz2016}, see the top row in Fig.~\ref{fig:exp_and_types}(b). And just as in the artificial spin ice, the degeneracy in the colloidal system is lifted due to the 2D square geometry, where all vertices are of type 3.

\begin{figure}[t]
\includegraphics[width=\columnwidth]{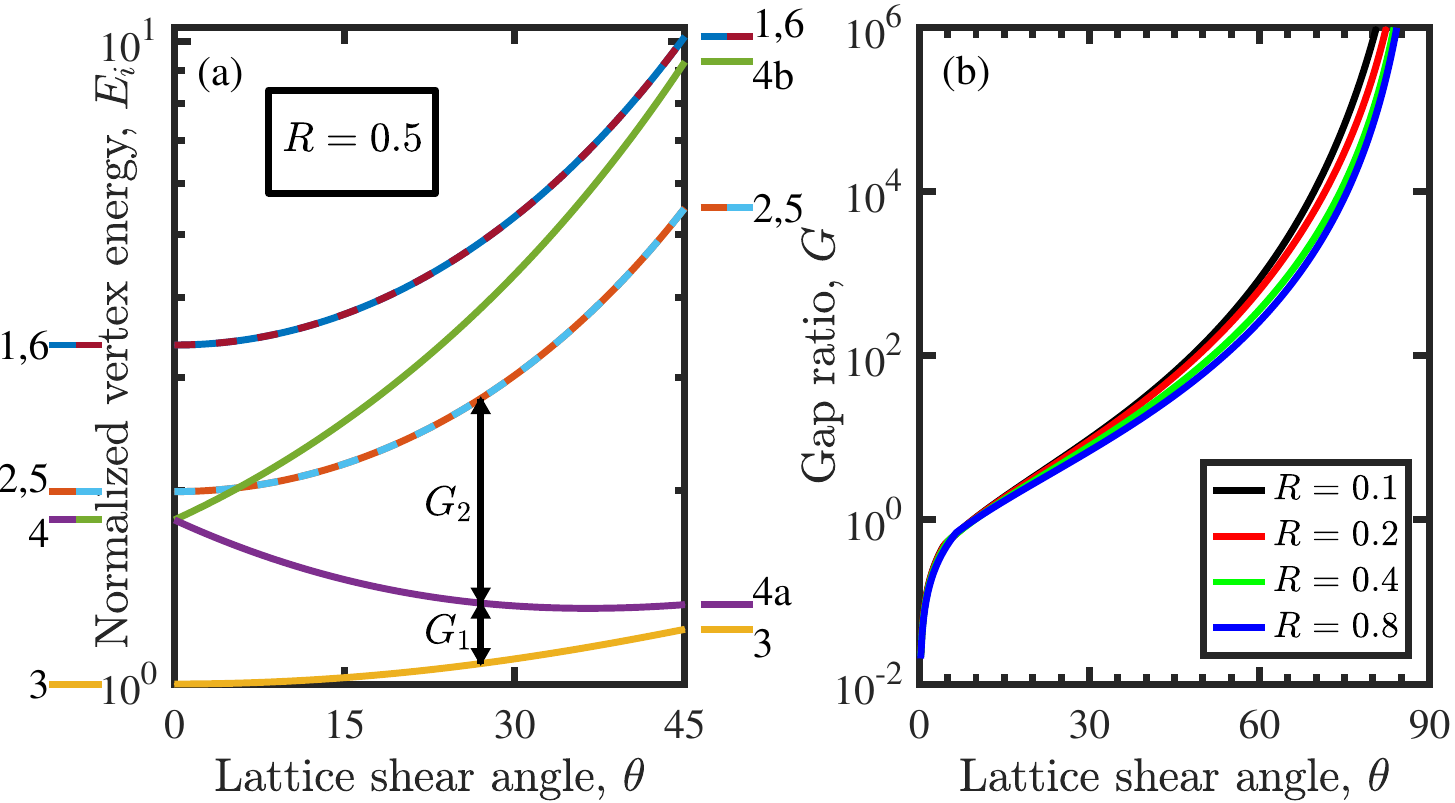}
\caption{(a) Evolution of the energy spectrum of the different types of a single vertex with increasing shear angle, ranging from the unsheared square lattice ($\theta=0$) on the left, to the maximally sheared lattice that we studied experimentally ($\theta=45^\circ$) on the right. Vertex types 2 and 5 and types 1 and 6 are clumped to particle-balanced effective vertices (2,5) and (1,6) with the average energies of these unbalanced types (see text).  (b) Ratio $G=G_2/G_1$ between the second and the first gaps in the energy spectrum depends weakly on the ratio $R=\ell/a$ of trap length to lattice constant, and reaches considerable values of 30-60 already at moderate shear of $\theta=45^\circ$.}
\label{fig:gap_ratio}
\end{figure}

In this Letter, we demonstrate an alternative approach to partially restore the ground-state degeneracy in square colloidal ice, by shearing the entire lattice by an angle $\theta$, as shown in our experimental demonstration in Fig.~\ref{fig:exp_and_types}(c). This strategy does not require to lift some of the double wells as suggested~\cite{Moller2006} and recently realized in artificial spin ice systems~\cite{Perrin2016}, thus it preserves the symmetry of magnetic dipolar interactions between the particles. As shown in Fig.~\ref{fig:exp_and_types}(b), the shearing leads to the splitting of the 2-in-2-out vertex type 4 into type 4a and type 4b, with type 4a having a lower energy than type 4b, see Fig. 2. With increasing shear angle, the energy of type 4a becomes arbitrarily close to that of type 3, as shown in Fig.~\ref{fig:gap_ratio}(a). If the energy gap $G_1$ between the type 3 and type 4a vertices is much smaller than the gap $G_2$ between type 4a and the next excited state of a vertex, then there exists an intermediate temperature range $G_1 \ll k_B T \ll G_2$, at which the occupancies of type 3 and type 4a vertices could be expected to be roughly equal and to dominate the system, thus an effective ground state could be obtained consisting of these two vertex types~\cite{Perrin2019}. We show that shearing the lattice can indeed lead to such extreme energy spectra with $G_1 \ll G_2$. However, the absence of type 4b vertices at low temperatures gives rise to topological differences between the low-energy configurations in this sheared square lattice and in the idealized square ice model~\cite{Lieb1967}, in which type 3 and all type 4 vertices have the same energy. Specifically, we show that even when type 4a vertices are energetically allowed, their occurrence requires higher-energy topological excitations involving type 2 and type 5 vertices.

We experimentally realize the colloidal ice by placing a suspension of paramagnetic colloidal particles of diameter $d=10 \mu m$ over a lithographically-patterned substrate of topographic double-well traps~\cite{Ortiz2016, Supplemental} arranged on the edges of a sheared square lattice. The trap length, defined as the distance between the two possible positions of the particle center in each double-well trap, is given by $\ell = 10\rm{\mu m}$, whereas the lattice constant is $a=36\rm{\mu m}$ for shear angle $\theta=25^\circ$, and $a=46\rm{\mu m}$ for $\theta=45^\circ$, see Fig.~\ref{fig:exp_and_types}(c). A perpendicular external magnetic field $\bm{B}$ induces in each particle of volume $V=\pi d^3/6$ a magnetic moment $\bm{m}=V \chi \bm{B}/\mu_0$, with $\mu_0 = 4 \pi 10^{-7} \rm{H/m}$ the vacuum permeability. We estimate the magnetic volume susceptibility of the particles as $\chi=0.017$. As a result, pairs of particles at a distance $r$ experience an isotropic repulsive interaction with energy $E = \mu_0 m^2/(4 \pi r^3)$. All experiments are at room temperature, but we gradually increase the external field $B$ from 0 to 8mT over a duration of 10 minutes, thus enabling us to control the strength of the interaction energy $E$ and by that to effectively change the thermal properties of the system without changing the temperature. Our data analysis includes particle tracking with labeling the vertex types in a lattice composed of 136 particles for $\theta=25^\circ$, and 115 for $45^\circ$, and performing an ensemble average over $10$ separate realizations for each shear angle.

We now outline our theoretical approach to describe the sheared system. Since between two vertices there is one particle in each double-well trap, and four traps meet at each vertex, the latter has on average two in-particles and two out-particles. Most generally, this requirement is obtained by requiring that the fractions $\{ p_i \}$ of vertices of the different \textit{unbalanced} types (i.e., vertex types with a number of in-particles different from 2) satisfy $p_1 \cdot 0 + p_2 \cdot 1 + p_5 \cdot 3 + p_6 \cdot 4 = 2$. Based on our experimental and numerical observations (see below), we assume that this global balancing of the number of in-particles per vertex is achieved by having $p_1=p_6$ and $p_2=p_5$. Namely, type 1 and type 6, and type 2 and type 5 vertices appear in (delocalized) pairs. Thus in order to calculate the occurrence of each vertex type, we will consider each $(1,6)$ pair to be constructed of two vertices, each with an effective energy $E_{1,6}=\left(E_1+E_6\right)/2$, and similarly each $(2,5)$ pair will be considered as two vertices, each with energy $E_{2,5}=\left(E_2+E_5\right)/2$.

Further, we assume negligible interactions for distances larger than the four particles meeting at each vertex. From the distances between these four particles, we directly calculate the contribution of each vertex to the total energy of the system. Since the interaction energy decays algebraically with distance, the ratios of the gaps between vertex energies depend only on the shear angle $\theta$ and on the dimensionless ratio $R = \ell / a$ of the trap length to the lattice constant. As shown in Fig.~\ref{fig:gap_ratio}(b), the resulting ratio $G=G_2/G_1$ between the second and the first energy gaps increases dramatically with shear angle, and depends  only weakly on the geometric parameter $R$. 

\begin{figure}[t]
\includegraphics[width=\columnwidth]{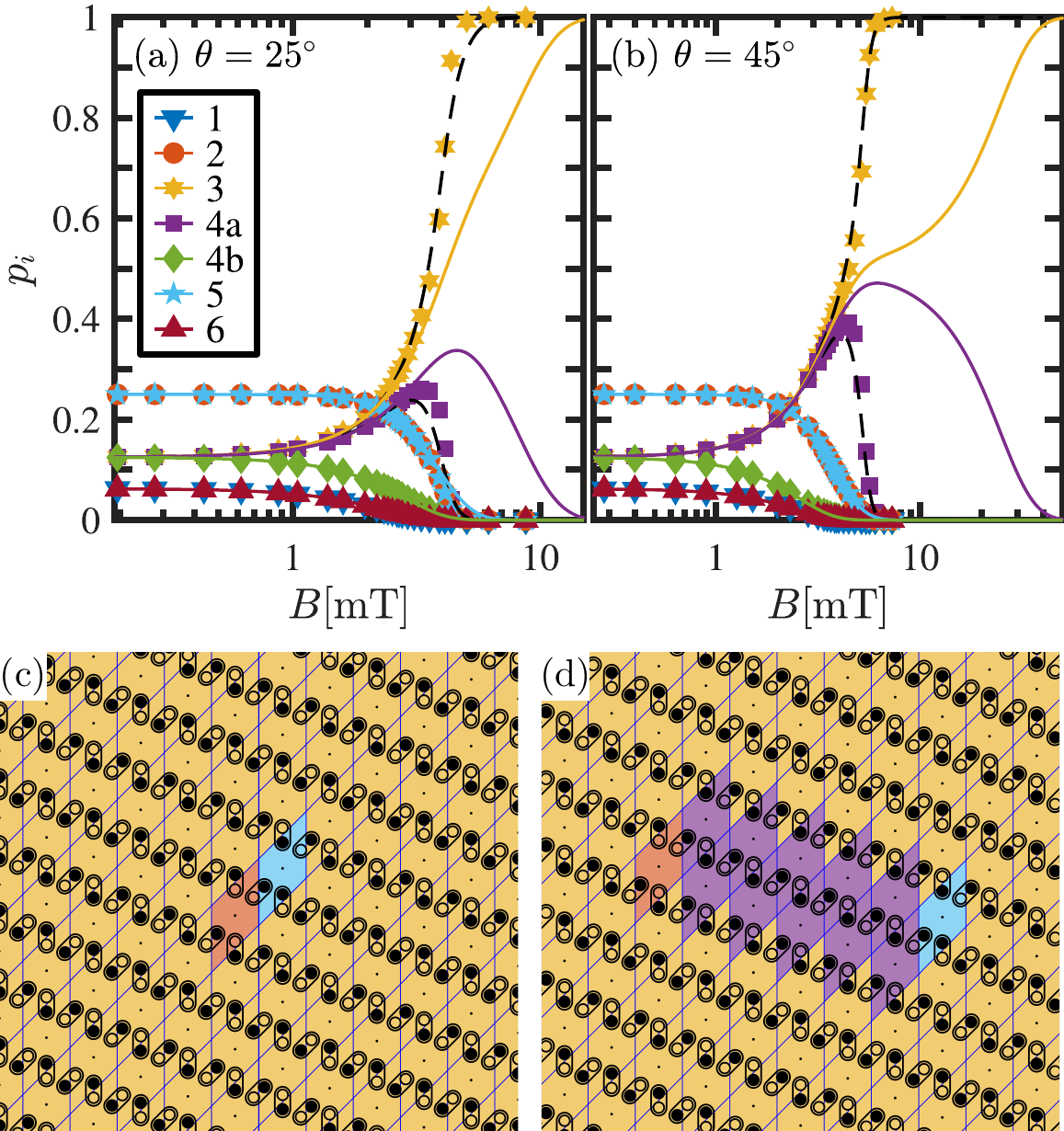}
\caption{Simulation results (symbols) vs. single-vertex mean-field theory (solid lines) and combined with 1D Ising theory for line defects (black dashed lines) for $\theta=25^\circ$(a), $45^\circ$(b). (c,d) Snapshots from simulation at $\theta=45^\circ$, showing how flipping the positions of a sequence of particles in the type 3 ground state generates a double row of type 4a vertices with type 2 and type 5 vertices surrounding this line defect. The Voronoi parallelogram around every vertex is colored according to the vertex type, following the color scheme in (a) and (b).}
\label{fig:line_defects}
\end{figure}

Assuming that each vertex in the lattice is statistically independent from the others, we may use the single-vertex effective energies $\{ E_{1,6} , E_{2,5} , E_3 , E_{4a} , E_{4b} , E_{2,5} , E_{1,6} \}$ and the combinatorial degeneracies $\{ g_i \} = \{ 1,4,2,2,2,4,1 \}$ of the different vertex types $i=\{1,2,3,4a,4b,5,6\}$ to write a single-vertex mean-field prediction $p_i = g_i \exp \left( - E_i / k_B T \right) / Z$ for the fractions of the different vertex types. Here $Z = \sum_i g_i \exp \left( - E_i / k_B T \right)$ is the canonical partition function. We ran Monte Carlo simulations with discrete positions of particles at the two ends of each trap and including interactions only with particles in traps meeting in each vertex. The simulations included 5000 particles with periodic boundary conditions and was run at each field value until equilibration, typically with up to $4 \times 10^7$ time steps. As shown in Fig.~\ref{fig:line_defects}, the mean-field approach fails at large applied field. In this situation it predicts a quasidegenerate ground state of type 3 and type 4a at intermediate fields (solid lines), in contrast to the simulation results (symbols) which shows a ground state filled by type 3. Note, however that the occupancies of the other vertex types are captured very well by this mean-field theory.

Type 4a and type 3 vertices cannot coexist without the presence of vertices of other types; starting from the type 3 ground state, the lowest energy excitation entails flipping the position of one particle in its trap. This generates a pair of type 2 and type 5 vertices, see Fig 3(c). Only once such a (2,5) pair is formed, vertices of type 4a may appear due to the flipping of additional particles. As shown in Fig.~\ref{fig:line_defects}(d), for strong magnetic fields, type 4a vertices appear in domains comprised of two rows of vertices and surrounded by a (2,5) pair. 

We theoretically describe the abundance of such line defects by mapping them to a 1D Ising model; for every pair of adjacent diagonal rows in the sheared lattice, each position along this double row could be either two type 3 vertices with energy $2E_3$ - a state we will denote as $\sigma_j=-1$ - or two type 4a vertices with energy $2E_{4a}$ - a state we will denote as $\sigma_j=+1$. Each interface between type 3 and type 4 domains along this double row involves a type 2 or type 5 vertex and thus costs an energy $E_{2,5}$. Hence, the energy of any sequence $\{ \sigma_j \}$ of type 3 and type 4 vertices along this double row is given by the Ising Hamiltonian $\mathcal{H} = - J \sum_j \sigma_j \sigma_{j+1} - h \sum_j \sigma_j$, where $h = E_{3} - E_{4a} = -G_1$ represents the difference in energy between type 3 and type 4a vertex pairs, and $J=(E_{2,5} - E_3)/2 = (G_1+G_2)/2$ represents the energetic cost of having a type 2 and type 5 pair at each interface between type 3 and type 4a. Using the exact solution of this 1D Ising model~\cite{Ising1925, SalinasBook}, we get its magnetization $\langle \sigma \rangle = p_{4a} - p_3$, from which we deduce the difference between the occupancies of type 4a and type 3 vertices. Further, we take the the sum $p_3+p_{4a}$ from the mean-field theory, since it well describes the occupancies of all other vertex types, and use the 1D Ising model for line defects only in order to get the difference $p_3-p_{4a}$. As shown by the black dashed lines in Fig.~\ref{fig:line_defects}(a,b), this combined analytical theory agrees exceptionally well with the simulations (symbols). Specifically, we see that type 2 and type 5 vertices are required in order to obtain substantial fractions of type 4a vertices, hence the topological need for type 2 and type 5 vertices prohibits the existence of a degenerate effective ground state comprised only of type 3 and type 4a vertices. In principle, a line of type 4a vertices can span the entire system without having any interfaces, which require type 2 and type 5 defects. However, such excitations will have lower energy than the line defects with finite length discussed above only in very small systems.

\begin{figure}[t]
\includegraphics[width=\columnwidth]{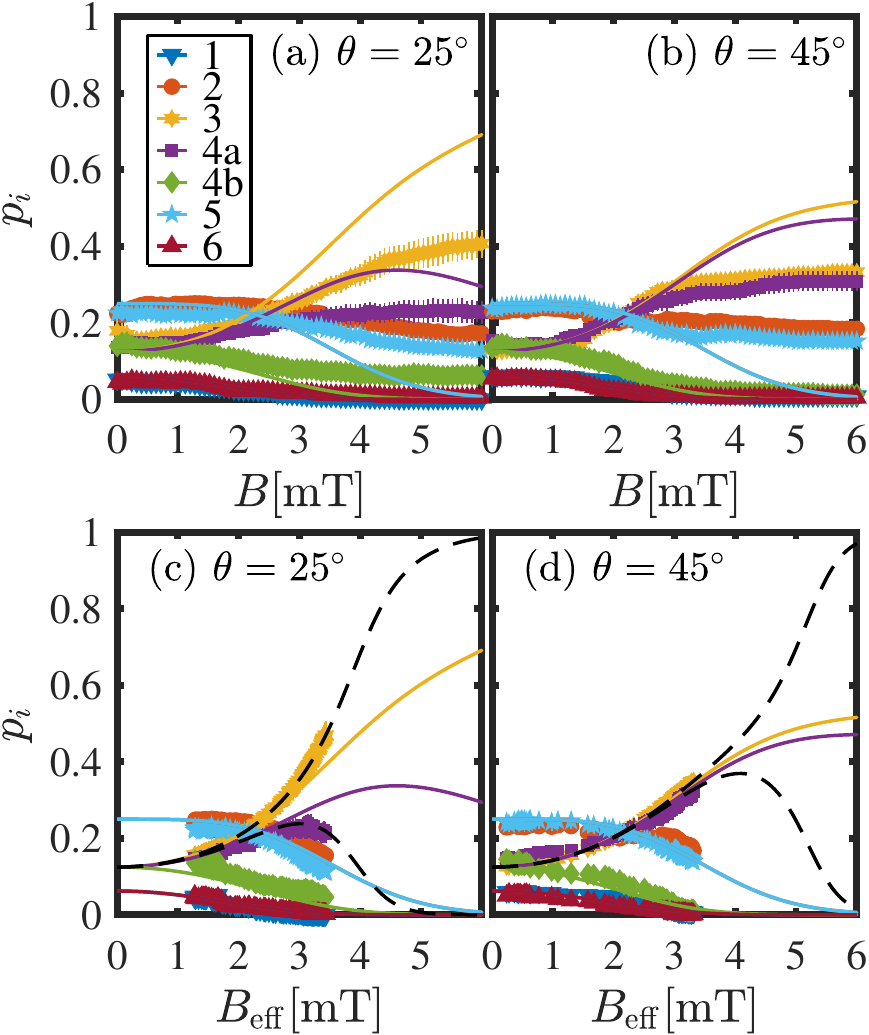}
\caption{Fractions of different types of vertices vs. magnetic field in experiment (symbols) and single-vertex mean-field theory (lines) for shear angles of $25^\circ$ (left) and $45^\circ$ (right). Results are plotted vs. the measured field (a,b) and vs. the effective field deduced from the occupancy of type 3 vertices (c,d). Black dashed lines show the exact solution of the 1D Ising model for our line defect theory.}
\label{fig:mf_exp}
\end{figure}

Figure~\ref{fig:mf_exp}(a,b) shows that our experiments and theory agree only for weak applied fields, where the interactions lead to small deviations from random occupations of the different vertex types. For field stronger than $\sim$3mT, the system  evolves much more slowly with increasing field than theoretically predicted. We suggest that at strong fields, the system falls into metastable states, where it requires much longer waiting times in order to equilibrate. The particles used are relatively large and their motion is damped by viscous dissipation. Therefore, they are rather slow in exploring the system's phase space. Together with the presence of disorder in the topographic traps, this allows some particles to relax before others and adds to the nonergodicity of the system. Similarly to other works on effective thermal descriptions for athermal systems~\cite{Cugliandolo1997, Ono2002, Nisoli2010, Cugliandolo2011}, we find that for any given field, the occupancies of the different vertex types may be described by an equilibrium-like Boltzmann distribution and using an effective magnetic field $B_{\mathrm{eff}}$ which is almost half the magnitude of the applied field. We use $p_3$ to extract $B_{\mathrm{eff}}$ at any measured field, and in Fig.~\ref{fig:mf_exp}(c,d) we plot the experimental data vs. $B_{\mathrm{eff}}$ rather than vs. $B$. This leads to an impressive agreement between experiment (symbols) and theory, including the deviation from mean field (solid lines) as explained by our Ising model to describe line defects (dashed black lines). Due to experimental limitation, we cannot increase the applied field above $8$mT ($B_{\mathrm{eff}}\sim 3$mT), while substantially higher fields are required in order to observe the peak in the occupation of type 4a vertices. An interesting direction could be testing the effect of the quench rate on this mapping to an effective field, particularly, in order to find, whether a slower quench can assist the system in equilibrating and may allow it to reach higher effective fields. Note that at $\theta=25^\circ$, although we prepare disordered samples at $B=0$, the occupancies of the different vertex types there do not precisely match their combinatorial weights, and we do not observe distributions representative of effective fields lower than $1.3$mT. This could be caused by additional interactions between particles in the absence of an external magnetic field, but we must also point out that for $\theta=45^\circ$ the effective field starts almost at zero. 

To conclude, we show that shearing a square colloidal ice by a defined angle allows to partially recover the ground state degeneracy of 3D water ice using a confined 2D structure. Our experimental findings are complemented with theory and numerical simulations, and by considering an effective field we obtain excellent agreement over all the range of parameters explored. Future open directions of our work include the use of slower field ramps or other annealing protocols to better equilibrate different regions of the system, or the use of additional in-plane field components to select different types of vertices. Moreover, the possibility of visualizing in situ the dynamics of colloids could enable further studies on how defects evolve and destroy 
the partially degenerate ground state of the sheared system. On the application side, the shearing technique introduced in this work could be easily extended to other artificially frustrated systems such as to artificial spin ice, ferromagnetic structures or other soft-condensed matter systems where it is possible to deform an underlying lattice structure. The prospect of engineering magnetic structures with multiple ground states could allow to easily select one type of the patterns in order to write and store logical information. This feature could be used to design a novel class of memories and devices based on motion of topological charges and defects~\cite{Mengotti2011,Libal2012,YWang2016,Loehr2016, YWang2018}. One could apply the same technique to other geometries, beyond  the square case, e.g. to the kagome lattice~\cite{Olson2012, Libal2018, Nisoli2018}, which will allow the exploration of many other physical situation with the associated complex and richer dynamics. 

We thank Ram Avinery, Andr\'as Lib\'al, Cristiano Nisoli, Tal Oz, Charles Reichhardt, and Yael Roichman for helpful discussions. This research was supported in part by the Israel Science Foundation Grant No. 968/16. Experiments were realized with the help of the MicroFabSpace and Microscopy Characterization Facility at IBEC. A. O.-A., and P. T. acknowledge support from the ERC Grant (Number 811234). P. T. acknowledges support from from MINECO (FIS2016-78507-C2, ERC2018-092827), DURSI (2017SGR1061), and Generalitat de Catalunya under Program ``ICREA Acadèmia''.

\bibliography{sheared_ice_bib}

\begin{thebibliography}{45}%
\makeatletter
\providecommand \@ifxundefined [1]{%
 \@ifx{#1\undefined}
}%
\providecommand \@ifnum [1]{%
 \ifnum #1\expandafter \@firstoftwo
 \else \expandafter \@secondoftwo
 \fi
}%
\providecommand \@ifx [1]{%
 \ifx #1\expandafter \@firstoftwo
 \else \expandafter \@secondoftwo
 \fi
}%
\providecommand \natexlab [1]{#1}%
\providecommand \enquote  [1]{``#1''}%
\providecommand \bibnamefont  [1]{#1}%
\providecommand \bibfnamefont [1]{#1}%
\providecommand \citenamefont [1]{#1}%
\providecommand \href@noop [0]{\@secondoftwo}%
\providecommand \href [0]{\begingroup \@sanitize@url \@href}%
\providecommand \@href[1]{\@@startlink{#1}\@@href}%
\providecommand \@@href[1]{\endgroup#1\@@endlink}%
\providecommand \@sanitize@url [0]{\catcode `\\12\catcode `\$12\catcode
  `\&12\catcode `\#12\catcode `\^12\catcode `\_12\catcode `\%12\relax}%
\providecommand \@@startlink[1]{}%
\providecommand \@@endlink[0]{}%
\providecommand \url  [0]{\begingroup\@sanitize@url \@url }%
\providecommand \@url [1]{\endgroup\@href {#1}{\urlprefix }}%
\providecommand \urlprefix  [0]{URL }%
\providecommand \Eprint [0]{\href }%
\providecommand \doibase [0]{http://dx.doi.org/}%
\providecommand \selectlanguage [0]{\@gobble}%
\providecommand \bibinfo  [0]{\@secondoftwo}%
\providecommand \bibfield  [0]{\@secondoftwo}%
\providecommand \translation [1]{[#1]}%
\providecommand \BibitemOpen [0]{}%
\providecommand \bibitemStop [0]{}%
\providecommand \bibitemNoStop [0]{.\EOS\space}%
\providecommand \EOS [0]{\spacefactor3000\relax}%
\providecommand \BibitemShut  [1]{\csname bibitem#1\endcsname}%
\let\auto@bib@innerbib\@empty
\bibitem [{\citenamefont {Pauling}(1935)}]{Pauling1935}%
  \BibitemOpen
  \bibfield  {author} {\bibinfo {author} {\bibfnamefont {L.}~\bibnamefont
  {Pauling}},\ }\href@noop {} {\bibfield  {journal} {\bibinfo  {journal} {J.
  Am. Chem. Soc.}\ }\textbf {\bibinfo {volume} {57}},\ \bibinfo {pages} {2680}
  (\bibinfo {year} {1935})}\BibitemShut {NoStop}%
\bibitem [{\citenamefont {Wang}\ \emph {et~al.}(2006)\citenamefont {Wang},
  \citenamefont {Nisoli}, \citenamefont {Freitas}, \citenamefont {Li},
  \citenamefont {McConville}, \citenamefont {Cooley}, \citenamefont {Lund},
  \citenamefont {Samarth}, \citenamefont {Leighton}, \citenamefont {Crespi},\
  and\ \citenamefont {Schiffer}}]{Wang2006}%
  \BibitemOpen
  \bibfield  {author} {\bibinfo {author} {\bibfnamefont {R.~F.}\ \bibnamefont
  {Wang}}, \bibinfo {author} {\bibfnamefont {C.}~\bibnamefont {Nisoli}},
  \bibinfo {author} {\bibfnamefont {R.~S.}\ \bibnamefont {Freitas}}, \bibinfo
  {author} {\bibfnamefont {J.}~\bibnamefont {Li}}, \bibinfo {author}
  {\bibfnamefont {W.}~\bibnamefont {McConville}}, \bibinfo {author}
  {\bibfnamefont {B.~J.}\ \bibnamefont {Cooley}}, \bibinfo {author}
  {\bibfnamefont {M.~S.}\ \bibnamefont {Lund}}, \bibinfo {author}
  {\bibfnamefont {N.}~\bibnamefont {Samarth}}, \bibinfo {author} {\bibfnamefont
  {C.}~\bibnamefont {Leighton}}, \bibinfo {author} {\bibfnamefont {V.~H.}\
  \bibnamefont {Crespi}}, \ and\ \bibinfo {author} {\bibfnamefont
  {P.}~\bibnamefont {Schiffer}},\ }\href@noop {} {\bibfield  {journal}
  {\bibinfo  {journal} {Nature}\ }\textbf {\bibinfo {volume} {439}},\ \bibinfo
  {pages} {303} (\bibinfo {year} {2006})}\BibitemShut {NoStop}%
\bibitem [{\citenamefont {Nisoli}\ \emph {et~al.}(2013)\citenamefont {Nisoli},
  \citenamefont {Moessner},\ and\ \citenamefont {Schiffer}}]{NisoliRMP2013}%
  \BibitemOpen
  \bibfield  {author} {\bibinfo {author} {\bibfnamefont {C.}~\bibnamefont
  {Nisoli}}, \bibinfo {author} {\bibfnamefont {R.}~\bibnamefont {Moessner}}, \
  and\ \bibinfo {author} {\bibfnamefont {P.}~\bibnamefont {Schiffer}},\
  }\href@noop {} {\bibfield  {journal} {\bibinfo  {journal} {Rev. Mod. Phys.}\
  }\textbf {\bibinfo {volume} {85}},\ \bibinfo {pages} {1473} (\bibinfo {year}
  {2013})}\BibitemShut {NoStop}%
\bibitem [{\citenamefont {Skj{\ae}rv{\o}}\ \emph {et~al.}(2019)\citenamefont
  {Skj{\ae}rv{\o}}, \citenamefont {Marrows}, \citenamefont {Stamps},\ and\
  \citenamefont {Heyderman}}]{Skjaervo2019}%
  \BibitemOpen
  \bibfield  {author} {\bibinfo {author} {\bibfnamefont {S.~H.}\ \bibnamefont
  {Skj{\ae}rv{\o}}}, \bibinfo {author} {\bibfnamefont {C.~H.}\ \bibnamefont
  {Marrows}}, \bibinfo {author} {\bibfnamefont {R.~L.}\ \bibnamefont {Stamps}},
  \ and\ \bibinfo {author} {\bibfnamefont {L.~J.}\ \bibnamefont {Heyderman}},\
  }\href@noop {} {\bibfield  {journal} {\bibinfo  {journal} {Nat. Rev. Phys.}\
  } (\bibinfo {year} {2019})}\BibitemShut {NoStop}%
\bibitem [{\citenamefont {Han}\ \emph {et~al.}(2008)\citenamefont {Han},
  \citenamefont {Shokef}, \citenamefont {Alsayed}, \citenamefont {Yunker},
  \citenamefont {Lubensky},\ and\ \citenamefont {Yodh}}]{Han2008}%
  \BibitemOpen
  \bibfield  {author} {\bibinfo {author} {\bibfnamefont {Y.}~\bibnamefont
  {Han}}, \bibinfo {author} {\bibfnamefont {Y.}~\bibnamefont {Shokef}},
  \bibinfo {author} {\bibfnamefont {A.~M.}\ \bibnamefont {Alsayed}}, \bibinfo
  {author} {\bibfnamefont {P.}~\bibnamefont {Yunker}}, \bibinfo {author}
  {\bibfnamefont {T.~C.}\ \bibnamefont {Lubensky}}, \ and\ \bibinfo {author}
  {\bibfnamefont {A.~G.}\ \bibnamefont {Yodh}},\ }\href@noop {} {\bibfield
  {journal} {\bibinfo  {journal} {Nature}\ }\textbf {\bibinfo {volume} {456}},\
  \bibinfo {pages} {898} (\bibinfo {year} {2008})}\BibitemShut {NoStop}%
\bibitem [{\citenamefont {Shokef}\ \emph {et~al.}(2011)\citenamefont {Shokef},
  \citenamefont {Souslov},\ and\ \citenamefont {Lubensky}}]{Shokef2011}%
  \BibitemOpen
  \bibfield  {author} {\bibinfo {author} {\bibfnamefont {Y.}~\bibnamefont
  {Shokef}}, \bibinfo {author} {\bibfnamefont {A.}~\bibnamefont {Souslov}}, \
  and\ \bibinfo {author} {\bibfnamefont {T.~C.}\ \bibnamefont {Lubensky}},\
  }\href@noop {} {\bibfield  {journal} {\bibinfo  {journal} {Proc. Natl. Acad.
  Sci.}\ }\textbf {\bibinfo {volume} {108}},\ \bibinfo {pages} {11804}
  (\bibinfo {year} {2011})}\BibitemShut {NoStop}%
\bibitem [{\citenamefont {Kang}\ \emph {et~al.}(2014)\citenamefont {Kang},
  \citenamefont {Shan}, \citenamefont {Ko{\v s}mrlj}, \citenamefont {Noorduin},
  \citenamefont {Shian}, \citenamefont {Weaver}, \citenamefont {Clark},\ and\
  \citenamefont {Bertoldi}}]{Kang2014}%
  \BibitemOpen
  \bibfield  {author} {\bibinfo {author} {\bibfnamefont {S.~H.}\ \bibnamefont
  {Kang}}, \bibinfo {author} {\bibfnamefont {S.}~\bibnamefont {Shan}}, \bibinfo
  {author} {\bibfnamefont {A.}~\bibnamefont {Ko{\v s}mrlj}}, \bibinfo {author}
  {\bibfnamefont {W.~L.}\ \bibnamefont {Noorduin}}, \bibinfo {author}
  {\bibfnamefont {S.}~\bibnamefont {Shian}}, \bibinfo {author} {\bibfnamefont
  {J.~C.}\ \bibnamefont {Weaver}}, \bibinfo {author} {\bibfnamefont {D.~R.}\
  \bibnamefont {Clark}}, \ and\ \bibinfo {author} {\bibfnamefont
  {K.}~\bibnamefont {Bertoldi}},\ }\href@noop {} {\bibfield  {journal}
  {\bibinfo  {journal} {Phys. Rev. Lett.}\ }\textbf {\bibinfo {volume} {112}},\
  \bibinfo {pages} {098701} (\bibinfo {year} {2014})}\BibitemShut {NoStop}%
\bibitem [{\citenamefont {Coulais}\ \emph {et~al.}(2017)\citenamefont
  {Coulais}, \citenamefont {Teomy}, \citenamefont {de~Reus}, \citenamefont
  {Shokef},\ and\ \citenamefont {van Hecke}}]{Coulais2016}%
  \BibitemOpen
  \bibfield  {author} {\bibinfo {author} {\bibfnamefont {C.}~\bibnamefont
  {Coulais}}, \bibinfo {author} {\bibfnamefont {E.}~\bibnamefont {Teomy}},
  \bibinfo {author} {\bibfnamefont {K.}~\bibnamefont {de~Reus}}, \bibinfo
  {author} {\bibfnamefont {Y.}~\bibnamefont {Shokef}}, \ and\ \bibinfo {author}
  {\bibfnamefont {M.}~\bibnamefont {van Hecke}},\ }\href@noop {} {\bibfield
  {journal} {\bibinfo  {journal} {Nature}\ }\textbf {\bibinfo {volume} {535}},\
  \bibinfo {pages} {529} (\bibinfo {year} {2017})}\BibitemShut {NoStop}%
\bibitem [{\citenamefont {Meeussen}\ \emph
  {et~al.}(2020{\natexlab{a}})\citenamefont {Meeussen}, \citenamefont {O{\u
  g}uz}, \citenamefont {Shokef},\ and\ \citenamefont {van
  Hecke}}]{Meeussen2019a}%
  \BibitemOpen
  \bibfield  {author} {\bibinfo {author} {\bibfnamefont {A.~S.}\ \bibnamefont
  {Meeussen}}, \bibinfo {author} {\bibfnamefont {E.~C.}\ \bibnamefont {O{\u
  g}uz}}, \bibinfo {author} {\bibfnamefont {Y.}~\bibnamefont {Shokef}}, \ and\
  \bibinfo {author} {\bibfnamefont {M.}~\bibnamefont {van Hecke}},\ }\href@noop
  {} {\bibfield  {journal} {\bibinfo  {journal} {Nat. Phys.
  DOI:10.1038/s41567-019-0763-6}\ } (\bibinfo {year}
  {2020}{\natexlab{a}})}\BibitemShut {NoStop}%
\bibitem [{\citenamefont {Meeussen}\ \emph
  {et~al.}(2020{\natexlab{b}})\citenamefont {Meeussen}, \citenamefont {O{\u
  g}uz}, \citenamefont {van Hecke},\ and\ \citenamefont
  {Shokef}}]{Meeussen2019b}%
  \BibitemOpen
  \bibfield  {author} {\bibinfo {author} {\bibfnamefont {A.~S.}\ \bibnamefont
  {Meeussen}}, \bibinfo {author} {\bibfnamefont {E.~C.}\ \bibnamefont {O{\u
  g}uz}}, \bibinfo {author} {\bibfnamefont {M.}~\bibnamefont {van Hecke}}, \
  and\ \bibinfo {author} {\bibfnamefont {Y.}~\bibnamefont {Shokef}},\
  }\href@noop {} {\bibfield  {journal} {\bibinfo  {journal} {New J. Phys.
  DOI:10.1088/1367-2630/ab69b5}\ } (\bibinfo {year}
  {2020}{\natexlab{b}})}\BibitemShut {NoStop}%
\bibitem [{\citenamefont {Lieb}(1967)}]{Lieb1967}%
  \BibitemOpen
  \bibfield  {author} {\bibinfo {author} {\bibfnamefont {E.~H.}\ \bibnamefont
  {Lieb}},\ }\href@noop {} {\bibfield  {journal} {\bibinfo  {journal} {Phys.
  Rev.}\ }\textbf {\bibinfo {volume} {162}},\ \bibinfo {pages} {162} (\bibinfo
  {year} {1967})}\BibitemShut {NoStop}%
\bibitem [{\citenamefont {Baez}\ and\ \citenamefont
  {Muniain}(1994)}]{BaezBook}%
  \BibitemOpen
  \bibfield  {author} {\bibinfo {author} {\bibfnamefont {J.}~\bibnamefont
  {Baez}}\ and\ \bibinfo {author} {\bibfnamefont {J.~P.}\ \bibnamefont
  {Muniain}},\ }\enquote {\bibinfo {title} {Gauge fields, knots and gravity},}\
  \ (\bibinfo  {publisher} {World Scientific},\ \bibinfo {address}
  {1stedition},\ \bibinfo {year} {1994})\BibitemShut {NoStop}%
\bibitem [{\citenamefont {Eckle}(2019)}]{EckleBook}%
  \BibitemOpen
  \bibfield  {author} {\bibinfo {author} {\bibfnamefont {H.-P.}\ \bibnamefont
  {Eckle}},\ }\enquote {\bibinfo {title} {Models of quantum matter: A first
  course on integrability and the bethe ansatz},}\ \ (\bibinfo  {publisher}
  {Oxford University Press},\ \bibinfo {address} {Oxford, UK},\ \bibinfo {year}
  {2019})\BibitemShut {NoStop}%
\bibitem [{\citenamefont {M{\"o}ller}\ and\ \citenamefont
  {Moessner}(2006)}]{Moller2006}%
  \BibitemOpen
  \bibfield  {author} {\bibinfo {author} {\bibfnamefont {G.}~\bibnamefont
  {M{\"o}ller}}\ and\ \bibinfo {author} {\bibfnamefont {R.}~\bibnamefont
  {Moessner}},\ }\href@noop {} {\bibfield  {journal} {\bibinfo  {journal}
  {Phys. Rev. Lett.}\ }\textbf {\bibinfo {volume} {96}},\ \bibinfo {pages}
  {237202} (\bibinfo {year} {2006})}\BibitemShut {NoStop}%
\bibitem [{\citenamefont {Perrin}\ \emph {et~al.}(2016)\citenamefont {Perrin},
  \citenamefont {Canals},\ and\ \citenamefont {Rougemaille}}]{Perrin2016}%
  \BibitemOpen
  \bibfield  {author} {\bibinfo {author} {\bibfnamefont {Y.}~\bibnamefont
  {Perrin}}, \bibinfo {author} {\bibfnamefont {B.}~\bibnamefont {Canals}}, \
  and\ \bibinfo {author} {\bibfnamefont {N.}~\bibnamefont {Rougemaille}},\
  }\href@noop {} {\bibfield  {journal} {\bibinfo  {journal} {Nature}\ }\textbf
  {\bibinfo {volume} {540}},\ \bibinfo {pages} {410} (\bibinfo {year}
  {2016})}\BibitemShut {NoStop}%
\bibitem [{\citenamefont {May}\ \emph {et~al.}(2019)\citenamefont {May},
  \citenamefont {Hunt}, \citenamefont {Berg}, \citenamefont {Hejazi},\ and\
  \citenamefont {Ladak}}]{May2019}%
  \BibitemOpen
  \bibfield  {author} {\bibinfo {author} {\bibfnamefont {A.}~\bibnamefont
  {May}}, \bibinfo {author} {\bibfnamefont {M.}~\bibnamefont {Hunt}}, \bibinfo
  {author} {\bibfnamefont {A.~V.~D.}\ \bibnamefont {Berg}}, \bibinfo {author}
  {\bibfnamefont {A.}~\bibnamefont {Hejazi}}, \ and\ \bibinfo {author}
  {\bibfnamefont {S.}~\bibnamefont {Ladak}},\ }\href@noop {} {\bibfield
  {journal} {\bibinfo  {journal} {Commun. Phys.}\ }\textbf {\bibinfo {volume}
  {2}},\ \bibinfo {pages} {13} (\bibinfo {year} {2019})}\BibitemShut {NoStop}%
\bibitem [{\citenamefont {{\"O}stman}\ \emph {et~al.}(2018)\citenamefont
  {{\"O}stman}, \citenamefont {Stopfel}, \citenamefont {Chioar}, \citenamefont
  {Arnalds}, \citenamefont {Stein}, \citenamefont {Kapaklis},\ and\
  \citenamefont {Hj{\"o}rvarsson}}]{Ostman2018}%
  \BibitemOpen
  \bibfield  {author} {\bibinfo {author} {\bibfnamefont {E.}~\bibnamefont
  {{\"O}stman}}, \bibinfo {author} {\bibfnamefont {H.}~\bibnamefont {Stopfel}},
  \bibinfo {author} {\bibfnamefont {I.-A.}\ \bibnamefont {Chioar}}, \bibinfo
  {author} {\bibfnamefont {U.~B.}\ \bibnamefont {Arnalds}}, \bibinfo {author}
  {\bibfnamefont {A.}~\bibnamefont {Stein}}, \bibinfo {author} {\bibfnamefont
  {V.}~\bibnamefont {Kapaklis}}, \ and\ \bibinfo {author} {\bibfnamefont
  {B.}~\bibnamefont {Hj{\"o}rvarsson}},\ }\href@noop {} {\bibfield  {journal}
  {\bibinfo  {journal} {Nat. Phys.}\ }\textbf {\bibinfo {volume} {14}},\
  \bibinfo {pages} {375} (\bibinfo {year} {2018})}\BibitemShut {NoStop}%
\bibitem [{\citenamefont {Perrin}\ \emph {et~al.}(2019)\citenamefont {Perrin},
  \citenamefont {Canals},\ and\ \citenamefont {Rougemaille}}]{Perrin2019}%
  \BibitemOpen
  \bibfield  {author} {\bibinfo {author} {\bibfnamefont {Y.}~\bibnamefont
  {Perrin}}, \bibinfo {author} {\bibfnamefont {B.}~\bibnamefont {Canals}}, \
  and\ \bibinfo {author} {\bibfnamefont {N.}~\bibnamefont {Rougemaille}},\
  }\href@noop {} {\bibfield  {journal} {\bibinfo  {journal} {Phys. Rev. B}\
  }\textbf {\bibinfo {volume} {99}},\ \bibinfo {pages} {224434} (\bibinfo
  {year} {2019})}\BibitemShut {NoStop}%
\bibitem [{\citenamefont {Caravelli}(2019)}]{Caravelli2019}%
  \BibitemOpen
  \bibfield  {author} {\bibinfo {author} {\bibfnamefont {F.}~\bibnamefont
  {Caravelli}},\ }\href@noop {} {\bibfield  {journal} {\bibinfo  {journal}
  {arXiv:1911.07028}\ } (\bibinfo {year} {2019})}\BibitemShut {NoStop}%
\bibitem [{\citenamefont {Tanaka}\ \emph {et~al.}(2006)\citenamefont {Tanaka},
  \citenamefont {Saitoh}, \citenamefont {Miyajima}, \citenamefont {Yamaoka},\
  and\ \citenamefont {Iye}}]{Tanaka2006}%
  \BibitemOpen
  \bibfield  {author} {\bibinfo {author} {\bibfnamefont {M.}~\bibnamefont
  {Tanaka}}, \bibinfo {author} {\bibfnamefont {E.}~\bibnamefont {Saitoh}},
  \bibinfo {author} {\bibfnamefont {H.}~\bibnamefont {Miyajima}}, \bibinfo
  {author} {\bibfnamefont {T.}~\bibnamefont {Yamaoka}}, \ and\ \bibinfo
  {author} {\bibfnamefont {Y.}~\bibnamefont {Iye}},\ }\href@noop {} {\bibfield
  {journal} {\bibinfo  {journal} {Phys. Rev. B}\ }\textbf {\bibinfo {volume}
  {73}},\ \bibinfo {pages} {052411} (\bibinfo {year} {2006})}\BibitemShut
  {NoStop}%
\bibitem [{\citenamefont {Morrison}\ \emph {et~al.}(2013)\citenamefont
  {Morrison}, \citenamefont {Nelson},\ and\ \citenamefont
  {Nisoli}}]{Morrison2013}%
  \BibitemOpen
  \bibfield  {author} {\bibinfo {author} {\bibfnamefont {M.~J.}\ \bibnamefont
  {Morrison}}, \bibinfo {author} {\bibfnamefont {T.~R.}\ \bibnamefont
  {Nelson}}, \ and\ \bibinfo {author} {\bibfnamefont {C.}~\bibnamefont
  {Nisoli}},\ }\href@noop {} {\bibfield  {journal} {\bibinfo  {journal} {New J.
  Phys.}\ }\textbf {\bibinfo {volume} {15}},\ \bibinfo {pages} {045009}
  (\bibinfo {year} {2013})}\BibitemShut {NoStop}%
\bibitem [{\citenamefont {Chern}\ \emph {et~al.}(2013)\citenamefont {Chern},
  \citenamefont {Morrison},\ and\ \citenamefont {Nisoli}}]{Chern2013}%
  \BibitemOpen
  \bibfield  {author} {\bibinfo {author} {\bibfnamefont {G.~W.}\ \bibnamefont
  {Chern}}, \bibinfo {author} {\bibfnamefont {M.~J.}\ \bibnamefont {Morrison}},
  \ and\ \bibinfo {author} {\bibfnamefont {C.}~\bibnamefont {Nisoli}},\
  }\href@noop {} {\bibfield  {journal} {\bibinfo  {journal} {Phys. Rev. Lett.}\
  }\textbf {\bibinfo {volume} {111}},\ \bibinfo {pages} {177201} (\bibinfo
  {year} {2013})}\BibitemShut {NoStop}%
\bibitem [{\citenamefont {Gilbert}\ \emph {et~al.}(2014)\citenamefont
  {Gilbert}, \citenamefont {Chern}, \citenamefont {Zhang}, \citenamefont
  {O'Brien}, \citenamefont {Fore}, \citenamefont {Nisoli},\ and\ \citenamefont
  {Schiffer}}]{Gilbert2014}%
  \BibitemOpen
  \bibfield  {author} {\bibinfo {author} {\bibfnamefont {I.}~\bibnamefont
  {Gilbert}}, \bibinfo {author} {\bibfnamefont {G.~W.}\ \bibnamefont {Chern}},
  \bibinfo {author} {\bibfnamefont {S.}~\bibnamefont {Zhang}}, \bibinfo
  {author} {\bibfnamefont {L.}~\bibnamefont {O'Brien}}, \bibinfo {author}
  {\bibfnamefont {B.}~\bibnamefont {Fore}}, \bibinfo {author} {\bibfnamefont
  {C.}~\bibnamefont {Nisoli}}, \ and\ \bibinfo {author} {\bibfnamefont
  {P.}~\bibnamefont {Schiffer}},\ }\href@noop {} {\bibfield  {journal}
  {\bibinfo  {journal} {Nat. Phys.}\ }\textbf {\bibinfo {volume} {10}},\
  \bibinfo {pages} {670} (\bibinfo {year} {2014})}\BibitemShut {NoStop}%
\bibitem [{\citenamefont {Mac\^edo}\ \emph {et~al.}(2018)\citenamefont
  {Mac\^edo}, \citenamefont {Macauley}, \citenamefont {Nascimento},\ and\
  \citenamefont {Stamps}}]{Macedo2018}%
  \BibitemOpen
  \bibfield  {author} {\bibinfo {author} {\bibfnamefont {R.}~\bibnamefont
  {Mac\^edo}}, \bibinfo {author} {\bibfnamefont {G.~M.}\ \bibnamefont
  {Macauley}}, \bibinfo {author} {\bibfnamefont {F.~S.}\ \bibnamefont
  {Nascimento}}, \ and\ \bibinfo {author} {\bibfnamefont {R.~L.}\ \bibnamefont
  {Stamps}},\ }\href@noop {} {\bibfield  {journal} {\bibinfo  {journal} {Phys.
  Rev. B}\ }\textbf {\bibinfo {volume} {98}},\ \bibinfo {pages} {014437}
  (\bibinfo {year} {2018})}\BibitemShut {NoStop}%
\bibitem [{\citenamefont {Macauley}\ \emph {et~al.}(2019)\citenamefont
  {Macauley}, \citenamefont {Paterson}, \citenamefont {Li}, \citenamefont
  {Mac\^edo}, \citenamefont {McVitie},\ and\ \citenamefont
  {Stamps}}]{Macauley2019}%
  \BibitemOpen
  \bibfield  {author} {\bibinfo {author} {\bibfnamefont {G.~M.}\ \bibnamefont
  {Macauley}}, \bibinfo {author} {\bibfnamefont {G.~W.}\ \bibnamefont
  {Paterson}}, \bibinfo {author} {\bibfnamefont {Y.}~\bibnamefont {Li}},
  \bibinfo {author} {\bibfnamefont {R.}~\bibnamefont {Mac\^edo}}, \bibinfo
  {author} {\bibfnamefont {S.}~\bibnamefont {McVitie}}, \ and\ \bibinfo
  {author} {\bibfnamefont {R.~L.}\ \bibnamefont {Stamps}},\ }\href@noop {}
  {\bibfield  {journal} {\bibinfo  {journal} {arXiv:1908.08903}\ } (\bibinfo
  {year} {2019})}\BibitemShut {NoStop}%
\bibitem [{\citenamefont {Lib\'{a}l}\ \emph {et~al.}(2006)\citenamefont
  {Lib\'{a}l}, \citenamefont {Reichhardt},\ and\ \citenamefont
  {Reichhardt}}]{Libal2006}%
  \BibitemOpen
  \bibfield  {author} {\bibinfo {author} {\bibfnamefont {A.}~\bibnamefont
  {Lib\'{a}l}}, \bibinfo {author} {\bibfnamefont {C.}~\bibnamefont
  {Reichhardt}}, \ and\ \bibinfo {author} {\bibfnamefont {C.~J.~O.}\
  \bibnamefont {Reichhardt}},\ }\href@noop {} {\bibfield  {journal} {\bibinfo
  {journal} {Phys. Rev. Lett.}\ }\textbf {\bibinfo {volume} {97}},\ \bibinfo
  {pages} {228302} (\bibinfo {year} {2006})}\BibitemShut {NoStop}%
\bibitem [{\citenamefont {Ortiz-Ambriz}\ \emph {et~al.}(2019)\citenamefont
  {Ortiz-Ambriz}, \citenamefont {Nisoli}, \citenamefont {Reichhardt},
  \citenamefont {Reichhardt},\ and\ \citenamefont {Tierno}}]{Ortiz2019}%
  \BibitemOpen
  \bibfield  {author} {\bibinfo {author} {\bibfnamefont {A.}~\bibnamefont
  {Ortiz-Ambriz}}, \bibinfo {author} {\bibfnamefont {C.}~\bibnamefont
  {Nisoli}}, \bibinfo {author} {\bibfnamefont {C.}~\bibnamefont {Reichhardt}},
  \bibinfo {author} {\bibfnamefont {C.~J.~O.}\ \bibnamefont {Reichhardt}}, \
  and\ \bibinfo {author} {\bibfnamefont {P.}~\bibnamefont {Tierno}},\
  }\href@noop {} {\bibfield  {journal} {\bibinfo  {journal} {Rev. Mod. Phys.}\
  }\textbf {\bibinfo {volume} {91}},\ \bibinfo {pages} {041003} (\bibinfo
  {year} {2019})}\BibitemShut {NoStop}%
\bibitem [{\citenamefont {Ortiz-Ambriz}\ and\ \citenamefont
  {Tierno}(2016)}]{Ortiz2016}%
  \BibitemOpen
  \bibfield  {author} {\bibinfo {author} {\bibfnamefont {A.}~\bibnamefont
  {Ortiz-Ambriz}}\ and\ \bibinfo {author} {\bibfnamefont {P.}~\bibnamefont
  {Tierno}},\ }\href@noop {} {\bibfield  {journal} {\bibinfo  {journal} {Nat.
  Commun.}\ ,\ \bibinfo {pages} {10575}} (\bibinfo {year} {2016})}\BibitemShut
  {NoStop}%
\bibitem [{Sup()}]{Supplemental}%
  \BibitemOpen
  \href@noop {} {}\bibinfo {howpublished} {See Supplemental Material for
  experimental details.}\BibitemShut {Stop}%
\bibitem [{\citenamefont {Ising}(1925)}]{Ising1925}%
  \BibitemOpen
  \bibfield  {author} {\bibinfo {author} {\bibfnamefont {E.}~\bibnamefont
  {Ising}},\ }\href@noop {} {\bibfield  {journal} {\bibinfo  {journal} {Z.
  Phys.}\ }\textbf {\bibinfo {volume} {31}},\ \bibinfo {pages} {253} (\bibinfo
  {year} {1925})}\BibitemShut {NoStop}%
\bibitem [{\citenamefont {Salinas}(2001)}]{SalinasBook}%
  \BibitemOpen
  \bibfield  {author} {\bibinfo {author} {\bibfnamefont {S.~R.~A.}\
  \bibnamefont {Salinas}},\ }\enquote {\bibinfo {title} {Introduction to
  statistical physics},}\ \ (\bibinfo  {publisher} {Springer},\ \bibinfo
  {address} {New York, NY},\ \bibinfo {year} {2001})\ Chap.~\bibinfo {chapter}
  {13}, pp.\ \bibinfo {pages} {257--262}\BibitemShut {NoStop}%
\bibitem [{\citenamefont {Cugliandolo}\ \emph {et~al.}(1997)\citenamefont
  {Cugliandolo}, \citenamefont {Kurchan},\ and\ \citenamefont
  {Peliti}}]{Cugliandolo1997}%
  \BibitemOpen
  \bibfield  {author} {\bibinfo {author} {\bibfnamefont {L.~F.}\ \bibnamefont
  {Cugliandolo}}, \bibinfo {author} {\bibfnamefont {J.}~\bibnamefont
  {Kurchan}}, \ and\ \bibinfo {author} {\bibfnamefont {L.}~\bibnamefont
  {Peliti}},\ }\href@noop {} {\bibfield  {journal} {\bibinfo  {journal} {Phys.
  Rev. E}\ }\textbf {\bibinfo {volume} {55}},\ \bibinfo {pages} {3898}
  (\bibinfo {year} {1997})}\BibitemShut {NoStop}%
\bibitem [{\citenamefont {Ono}\ \emph {et~al.}(2002)\citenamefont {Ono},
  \citenamefont {O'Hern}, \citenamefont {Durian}, \citenamefont {Langer},
  \citenamefont {Liu},\ and\ \citenamefont {Nagel}}]{Ono2002}%
  \BibitemOpen
  \bibfield  {author} {\bibinfo {author} {\bibfnamefont {I.~K.}\ \bibnamefont
  {Ono}}, \bibinfo {author} {\bibfnamefont {C.~S.}\ \bibnamefont {O'Hern}},
  \bibinfo {author} {\bibfnamefont {D.~J.}\ \bibnamefont {Durian}}, \bibinfo
  {author} {\bibfnamefont {S.~A.}\ \bibnamefont {Langer}}, \bibinfo {author}
  {\bibfnamefont {A.~J.}\ \bibnamefont {Liu}}, \ and\ \bibinfo {author}
  {\bibfnamefont {S.~R.}\ \bibnamefont {Nagel}},\ }\href@noop {} {\bibfield
  {journal} {\bibinfo  {journal} {Phys. Rev. Lett.}\ }\textbf {\bibinfo
  {volume} {89}},\ \bibinfo {pages} {095703} (\bibinfo {year}
  {2002})}\BibitemShut {NoStop}%
\bibitem [{\citenamefont {Nisoli}\ \emph {et~al.}(2010)\citenamefont {Nisoli},
  \citenamefont {Li}, \citenamefont {Ke}, \citenamefont {Garand}, \citenamefont
  {Schiffer},\ and\ \citenamefont {Crespi}}]{Nisoli2010}%
  \BibitemOpen
  \bibfield  {author} {\bibinfo {author} {\bibfnamefont {C.}~\bibnamefont
  {Nisoli}}, \bibinfo {author} {\bibfnamefont {J.}~\bibnamefont {Li}}, \bibinfo
  {author} {\bibfnamefont {X.}~\bibnamefont {Ke}}, \bibinfo {author}
  {\bibfnamefont {D.}~\bibnamefont {Garand}}, \bibinfo {author} {\bibfnamefont
  {P.}~\bibnamefont {Schiffer}}, \ and\ \bibinfo {author} {\bibfnamefont
  {V.~H.}\ \bibnamefont {Crespi}},\ }\href@noop {} {\bibfield  {journal}
  {\bibinfo  {journal} {Phys. Rev. Lett.}\ }\textbf {\bibinfo {volume} {105}},\
  \bibinfo {pages} {047205} (\bibinfo {year} {2010})}\BibitemShut {NoStop}%
\bibitem [{\citenamefont {Cugliandolo}(2011)}]{Cugliandolo2011}%
  \BibitemOpen
  \bibfield  {author} {\bibinfo {author} {\bibfnamefont {L.~F.}\ \bibnamefont
  {Cugliandolo}},\ }\href@noop {} {\bibfield  {journal} {\bibinfo  {journal}
  {J. Phys. A: Math. Theor.}\ }\textbf {\bibinfo {volume} {44}},\ \bibinfo
  {pages} {483001} (\bibinfo {year} {2011})}\BibitemShut {NoStop}%
\bibitem [{\citenamefont {Mengotti}\ \emph {et~al.}(2011)\citenamefont
  {Mengotti}, \citenamefont {Heyderman}, \citenamefont {Rodr\'iguez},
  \citenamefont {Nolting}, \citenamefont {H\"ugli},\ and\ \citenamefont
  {Braun}}]{Mengotti2011}%
  \BibitemOpen
  \bibfield  {author} {\bibinfo {author} {\bibfnamefont {E.}~\bibnamefont
  {Mengotti}}, \bibinfo {author} {\bibfnamefont {L.~J.}\ \bibnamefont
  {Heyderman}}, \bibinfo {author} {\bibfnamefont {A.~F.}\ \bibnamefont
  {Rodr\'iguez}}, \bibinfo {author} {\bibfnamefont {F.}~\bibnamefont
  {Nolting}}, \bibinfo {author} {\bibfnamefont {R.~V.}\ \bibnamefont
  {H\"ugli}}, \ and\ \bibinfo {author} {\bibfnamefont {H.-B.}\ \bibnamefont
  {Braun}},\ }\href {https://doi.org/10.1038/nphys1794} {\bibfield  {journal}
  {\bibinfo  {journal} {Nat. Phys.}\ }\textbf {\bibinfo {volume} {7}},\
  \bibinfo {pages} {68–74} (\bibinfo {year} {2011})}\BibitemShut {NoStop}%
\bibitem [{\citenamefont {Lib\'al}\ \emph {et~al.}(2012)\citenamefont
  {Lib\'al}, \citenamefont {Reichhardt},\ and\ \citenamefont
  {Olson~Reichhardt}}]{Libal2012}%
  \BibitemOpen
  \bibfield  {author} {\bibinfo {author} {\bibfnamefont {A.}~\bibnamefont
  {Lib\'al}}, \bibinfo {author} {\bibfnamefont {C.}~\bibnamefont {Reichhardt}},
  \ and\ \bibinfo {author} {\bibfnamefont {C.~J.}\ \bibnamefont
  {Olson~Reichhardt}},\ }\href {\doibase 210.1103/PhysRevE.86.021406}
  {\bibfield  {journal} {\bibinfo  {journal} {Phys. Rev. E}\ }\textbf {\bibinfo
  {volume} {86}},\ \bibinfo {pages} {021406} (\bibinfo {year}
  {2012})}\BibitemShut {NoStop}%
\bibitem [{\citenamefont {Wang}\ \emph {et~al.}(2016)\citenamefont {Wang},
  \citenamefont {Xiao}, \citenamefont {Snezhko}, \citenamefont {Xu},
  \citenamefont {Ocola}, \citenamefont {Divan}, \citenamefont {Pearson},
  \citenamefont {Crabtree},\ and\ \citenamefont {Kwok}}]{YWang2016}%
  \BibitemOpen
  \bibfield  {author} {\bibinfo {author} {\bibfnamefont {Y.-L.}\ \bibnamefont
  {Wang}}, \bibinfo {author} {\bibfnamefont {Z.-L.}\ \bibnamefont {Xiao}},
  \bibinfo {author} {\bibfnamefont {A.}~\bibnamefont {Snezhko}}, \bibinfo
  {author} {\bibfnamefont {J.}~\bibnamefont {Xu}}, \bibinfo {author}
  {\bibfnamefont {L.~E.}\ \bibnamefont {Ocola}}, \bibinfo {author}
  {\bibfnamefont {R.}~\bibnamefont {Divan}}, \bibinfo {author} {\bibfnamefont
  {J.~E.}\ \bibnamefont {Pearson}}, \bibinfo {author} {\bibfnamefont {G.~W.}\
  \bibnamefont {Crabtree}}, \ and\ \bibinfo {author} {\bibfnamefont {W.-K.}\
  \bibnamefont {Kwok}},\ }\href {\doibase 10.1126/science.aad8037} {\bibfield
  {journal} {\bibinfo  {journal} {Science}\ }\textbf {\bibinfo {volume}
  {352}},\ \bibinfo {pages} {962} (\bibinfo {year} {2016})}\BibitemShut
  {NoStop}%
\bibitem [{\citenamefont {Loehr}\ \emph {et~al.}(2016)\citenamefont {Loehr},
  \citenamefont {Ortiz-Ambriz},\ and\ \citenamefont {Tierno}}]{Loehr2016}%
  \BibitemOpen
  \bibfield  {author} {\bibinfo {author} {\bibfnamefont {J.}~\bibnamefont
  {Loehr}}, \bibinfo {author} {\bibfnamefont {A.}~\bibnamefont {Ortiz-Ambriz}},
  \ and\ \bibinfo {author} {\bibfnamefont {P.}~\bibnamefont {Tierno}},\
  }\href@noop {} {\bibfield  {journal} {\bibinfo  {journal} {Phys. Rev. Lett.}\
  }\textbf {\bibinfo {volume} {117}},\ \bibinfo {pages} {168001} (\bibinfo
  {year} {2016})}\BibitemShut {NoStop}%
\bibitem [{\citenamefont {Wang}\ \emph {et~al.}(2018)\citenamefont {Wang},
  \citenamefont {Ma}, \citenamefont {Xu}, \citenamefont {Xiao}, \citenamefont
  {Snezhko}, \citenamefont {Divan}, \citenamefont {Ocola}, \citenamefont
  {Pearson}, \citenamefont {Janko},\ and\ \citenamefont {Kwok}}]{YWang2018}%
  \BibitemOpen
  \bibfield  {author} {\bibinfo {author} {\bibfnamefont {Y.-L.}\ \bibnamefont
  {Wang}}, \bibinfo {author} {\bibfnamefont {X.}~\bibnamefont {Ma}}, \bibinfo
  {author} {\bibfnamefont {J.}~\bibnamefont {Xu}}, \bibinfo {author}
  {\bibfnamefont {Z.-L.}\ \bibnamefont {Xiao}}, \bibinfo {author}
  {\bibfnamefont {A.}~\bibnamefont {Snezhko}}, \bibinfo {author} {\bibfnamefont
  {R.}~\bibnamefont {Divan}}, \bibinfo {author} {\bibfnamefont {L.~E.}\
  \bibnamefont {Ocola}}, \bibinfo {author} {\bibfnamefont {J.~E.}\ \bibnamefont
  {Pearson}}, \bibinfo {author} {\bibfnamefont {B.}~\bibnamefont {Janko}}, \
  and\ \bibinfo {author} {\bibfnamefont {W.-K.}\ \bibnamefont {Kwok}},\ }\href
  {https://doi.org/10.1038/s41565-018-0162-7} {\bibfield  {journal} {\bibinfo
  {journal} {Nat. Nanothechnology}\ }\textbf {\bibinfo {volume} {13}},\
  \bibinfo {pages} {560} (\bibinfo {year} {2018})}\BibitemShut {NoStop}%
\bibitem [{\citenamefont {Reichhardt}\ \emph {et~al.}(2012)\citenamefont
  {Reichhardt}, \citenamefont {Lib\'{a}l},\ and\ \citenamefont
  {Reichhardt}}]{Olson2012}%
  \BibitemOpen
  \bibfield  {author} {\bibinfo {author} {\bibfnamefont {C.~J.~O.}\
  \bibnamefont {Reichhardt}}, \bibinfo {author} {\bibfnamefont
  {A.}~\bibnamefont {Lib\'{a}l}}, \ and\ \bibinfo {author} {\bibfnamefont
  {C.}~\bibnamefont {Reichhardt}},\ }\href@noop {} {\bibfield  {journal}
  {\bibinfo  {journal} {New J. Phys.}\ }\textbf {\bibinfo {volume} {14}},\
  \bibinfo {pages} {025006} (\bibinfo {year} {2012})}\BibitemShut {NoStop}%
\bibitem [{\citenamefont {Lib\'{a}l}\ \emph {et~al.}(2018)\citenamefont
  {Lib\'{a}l}, \citenamefont {Nisoli}, \citenamefont {Reichhardt},\ and\
  \citenamefont {Reichhardt}}]{Libal2018}%
  \BibitemOpen
  \bibfield  {author} {\bibinfo {author} {\bibfnamefont {A.}~\bibnamefont
  {Lib\'{a}l}}, \bibinfo {author} {\bibfnamefont {C.}~\bibnamefont {Nisoli}},
  \bibinfo {author} {\bibfnamefont {C.~J.~O.}\ \bibnamefont {Reichhardt}}, \
  and\ \bibinfo {author} {\bibfnamefont {C.}~\bibnamefont {Reichhardt}},\
  }\href@noop {} {\bibfield  {journal} {\bibinfo  {journal} {Phys. Rev. Lett.}\
  }\textbf {\bibinfo {volume} {120}},\ \bibinfo {pages} {027204} (\bibinfo
  {year} {2018})}\BibitemShut {NoStop}%
\bibitem [{\citenamefont {Nisoli}(2018)}]{Nisoli2018}%
  \BibitemOpen
  \bibfield  {author} {\bibinfo {author} {\bibfnamefont {C.}~\bibnamefont
  {Nisoli}},\ }\href@noop {} {\bibfield  {journal} {\bibinfo  {journal} {Phys.
  Rev. Lett.}\ }\textbf {\bibinfo {volume} {120}},\ \bibinfo {pages} {167205}
  (\bibinfo {year} {2018})}\BibitemShut {NoStop}%
\bibitem [{\citenamefont {Allan}\ \emph {et~al.}(2019)\citenamefont {Allan},
  \citenamefont {Wel}, \citenamefont {Keim}, \citenamefont {Caswell},
  \citenamefont {Wieker}, \citenamefont {Verweij}, \citenamefont {Reid},
  \citenamefont {{Thierry}}, \citenamefont {Grueter}, \citenamefont {Ramos},
  \citenamefont {{apiszcz}}, \citenamefont {{zoeith}}, \citenamefont {Perry},
  \citenamefont {Boulogne}, \citenamefont {Sinha}, \citenamefont
  {{pfigliozzi}}, \citenamefont {Bruot}, \citenamefont {Uieda}, \citenamefont
  {Katins}, \citenamefont {Mary},\ and\ \citenamefont {Ahmadia}}]{Trackpy}%
  \BibitemOpen
  \bibfield  {author} {\bibinfo {author} {\bibfnamefont {D.}~\bibnamefont
  {Allan}}, \bibinfo {author} {\bibfnamefont {C.~v.~d.}\ \bibnamefont {Wel}},
  \bibinfo {author} {\bibfnamefont {N.}~\bibnamefont {Keim}}, \bibinfo {author}
  {\bibfnamefont {T.~A.}\ \bibnamefont {Caswell}}, \bibinfo {author}
  {\bibfnamefont {D.}~\bibnamefont {Wieker}}, \bibinfo {author} {\bibfnamefont
  {R.}~\bibnamefont {Verweij}}, \bibinfo {author} {\bibfnamefont
  {C.}~\bibnamefont {Reid}}, \bibinfo {author} {\bibnamefont {{Thierry}}},
  \bibinfo {author} {\bibfnamefont {L.}~\bibnamefont {Grueter}}, \bibinfo
  {author} {\bibfnamefont {K.}~\bibnamefont {Ramos}}, \bibinfo {author}
  {\bibnamefont {{apiszcz}}}, \bibinfo {author} {\bibnamefont {{zoeith}}},
  \bibinfo {author} {\bibfnamefont {R.~W.}\ \bibnamefont {Perry}}, \bibinfo
  {author} {\bibfnamefont {F.}~\bibnamefont {Boulogne}}, \bibinfo {author}
  {\bibfnamefont {P.}~\bibnamefont {Sinha}}, \bibinfo {author} {\bibnamefont
  {{pfigliozzi}}}, \bibinfo {author} {\bibfnamefont {N.}~\bibnamefont {Bruot}},
  \bibinfo {author} {\bibfnamefont {L.}~\bibnamefont {Uieda}}, \bibinfo
  {author} {\bibfnamefont {J.}~\bibnamefont {Katins}}, \bibinfo {author}
  {\bibfnamefont {H.}~\bibnamefont {Mary}}, \ and\ \bibinfo {author}
  {\bibfnamefont {A.}~\bibnamefont {Ahmadia}},\ }\href {\doibase
  10.5281/ZENODO.3492186} {\enquote {\bibinfo {title} {{soft-matter/trackpy:
  Trackpy v0.4.2}},}\ } (\bibinfo {year} {2019})\BibitemShut {NoStop}%
\bibitem [{\citenamefont {Crocker}\ and\ \citenamefont
  {Grier}(1996)}]{Crocker1996e}%
  \BibitemOpen
  \bibfield  {author} {\bibinfo {author} {\bibfnamefont {J.~C.}\ \bibnamefont
  {Crocker}}\ and\ \bibinfo {author} {\bibfnamefont {D.~G.}\ \bibnamefont
  {Grier}},\ }\href@noop {} {\bibfield  {journal} {\bibinfo  {journal} {J.
  Colloid Interf. Sci.}\ }\textbf {\bibinfo {volume} {179}},\ \bibinfo {pages}
  {298} (\bibinfo {year} {1996})}\BibitemShut {NoStop}%
\end{thebibliography}%

\newpage

\section{Supplemental Material}

\subsection{Structure Lithography}

We realize topographical traps following the soft lithography method outlined in \cite{Ortiz2016}. A $Cr$ mask is fabricated by Direct Write Lithography (DWL66, Heidelberg Instruments Mikrotechnik GmbH), with a $\lambda=405$nm diode laser at a $5.7$mm$^2$/min writing speed. Then we spin coat a coverslip (No. 1, $\sim 100\mu$m Thermo Fisher) with a $2.5\mu$m layer of photoresist (AZ1512HS, Microchemicals). For this, we clean coverglasses with soap and water, acetone, and isopropanol (in that order) and then we dry them thoroughly with $N_2$ and by placing them in a hotplate at $120^\circ$C for $15$ minutes. We then spin the photoresist for 30 seconds at 1000 r.p.m. before curing it for 3 minutes at $95^\circ$C. After that we etch the structures by exposing selectively through the Cr mask with UV light at $21$mW/cm$^2$ (UV-NIL, SUSS Microtech) for 3.4 seconds and then developing the exposed parts by dipping the sample for 7 seconds in a developer solution (AZ726MIF). 

\subsection{Sample Preparation}

We prepare a colloidal suspension by diluting a stock suspension of  $10 \mu$m polystyrene particles doped with iron oxide (49664 Sigma-Aldrich) on a solution of ultrapure water (Direct-Q, Merk Millipore) and sodium dodecyl sulfate (SDS, Sigma-Aldrich) at 0.5 the critical micelle concentration. The SDS is used to prevent the adhesion of particles to the photoresist. We also added a small amount of tetramethyl amonium hydroxide (TMAH, Sigma-Aldrich) to the solution until the measured pH is 7. Before assembling the sample, the suspension was mixed for two days on a rotator (Mini LabRoller) inside a refrigerator, as we found this step reduces adhesion further. A droplet of the suspension is then sandwiched between the lithography substrate and a second coverslip and sealed with vacuum grease (Thorlabs). 

\subsection{Colloidal Ice Assembly}

The sample is placed on an inverted microscope (TiU, Nikon) fitted with an oil immersion 40x objective (NA = 1.4, Nikon) and a long-pass color filter (FEL0500, Thorlabs) to prevent the sample from degrading under the UV components of the incandescent illumination. When the sample is placed on the microscope, particles sediment randomly at the bottom of the sample due to density mismatch. We use optical tweezers to place one particle in every topographic double-well trap within the field of view, which yields a system of 8$\times$11 vertical traps and 8x11 slanted traps for $\theta=25^\circ$ and 7$\times$11 for $\theta=45^\circ$. In the analysis we take a subset of 6$\times$10 and 5$\times$10 full vertices to avoid the effect of vertices partially out of the field of view. 

The optical tweezers are realized by expanding the beam of a butterfly laser diode ($\lambda=976$nm, 300mW, operated at 70mW, BL976-SAG300 Thorlabs), introducing it through the epi-illumination module of the microscope, through a dichroic mirror (FF825-SD01-25x36x2.0, Semrock) and into the observation objective. Due to the magnetic domains, the colloidal particles can absorb the focused light of the laser, and create convective flows that destabilize the trap. This is normally addressed by keeping the laser power very low. Trap stability can be improved slightly by using a ring trap created with a spatial light modulator (SLM, Hamamatsu X10468-03). The SLM is conjugated with the back focal plane of the objective, and the ring trap is created by projecting the phase of a Bessel beam of $k_t = 50$mm$^{-1}$, $m=10$ and $l=0$. 

We generate an external magnetic field $\bm{B}$ using a pair of custom made Helmholtz coils, connected to a power amplifier (BOP-20 10M, KEPCO), which is controlled through a digital analogue card (NI 9269) and a custom made LabVIEW program. The coils are placed with their axis normal to the sample plane, so the only component of the field is the vertical component. Once the colloidal ice has been assembled, the experiments are performed by randomizing the initial configuration using a digital dice and flipping every particle with a 50\% probability. We then ramp up the magnetic field from 0 to 8mT during an interval of 10 min, while recording a video using a digital camera (MQ013CG-E2, Ximea) at 30fps. The positions of the particles are tracked using the Trackpy implementation \cite{Trackpy} of the Crocker-Grier algorithm \cite{Crocker1996e}. From the particle tracking we can extract the vertex types, and the magnetic field is connected to the time of the video using both series' timestamp. While counting the vertex types, we ignore the vertices at the open boundaries. 

\end{document}